\documentclass[preprint]{aastex62}

\shorttitle{Young, Spectroscopic Binaries}
\shortauthors{Flagg et al.}

\begin{document}


\title{ACRONYM IV: Three New, Young, Low-mass Spectroscopic Binaries}


\author{Laura Flagg}
\affil{Department of Physics and Astronomy, Rice University, 6100 Main St. MS-108, Houston, TX 77005, USA}
\affil{Department of Physics and Astronomy, Northern Arizona University, Box 6010, Flagstaff, AZ 86011, USA}
\affil{Lowell Observatory, 1400 W Mars Hill Road, Flagstaff, AZ 86001, USA}
\email{laura.flagg@rice.edu}

\author{Evgenya L. Shkolnik}
\affil{School of Earth and Space Exploration, Arizona State University, 781 S Terrace Road, Tempe, AZ 85281, USA}

\author{Alycia Weinberger}
\affil{Department of Terrestrial Magnetism, Carnegie Institution for Science, 5241 Broad Branch Road, NW, Washington, DC 20015, USA}

\author{Brendan P. Bowler}
\affil{McDonald Observatory and the Department of Astronomy, The University of Texas at Austin, Austin, TX 78712, USA}

\author{Brian Skiff}
\affil{Lowell Observatory, 1400 W Mars Hill Road, Flagstaff, AZ 86001, USA}

\author{Adam L. Kraus}
\affil{McDonald Observatory and the Department of Astronomy, The University of Texas at Austin, Austin, TX 78712, USA}

\author{Michael C. Liu}
\affil{Institute for Astronomy, University of Hawai\'i at M$\hat{a}$noa, 2680 Woodlawn Drive, Honolulu, HI 96822, USA}
%
%
%
%

\begin{abstract}

As part of our search for new low-mass members of nearby young moving groups (YMG), we discovered three low-mass, spectroscopic binaries, two of which are not kinematically associated with any known YMG. Using high-resolution optical spectroscopy, we measure the component and systemic radial velocities of the systems, as well as their lithium absorption and H$\alpha$ emission, both spectroscopic indicators of youth. One system (2MASS J02543316-5108313, M2.0+M3.0) we confirm as a member of the 40 Myr old Tuc-Hor moving group, but whose binarity was previously undetected.  The second young binary (2MASS J08355977-3042306, K5.5+M1.5) is not a kinematic match to any known YMG, but each component exhibits lithium absorption and strong and wide H$\alpha$ emission indicative of active accretion, setting an upper age limit of 15 Myr.  The third system (2MASS J10260210-4105537, M1.0+M3.0) has been hypothesized in the literature to be a member of the 10 Myr old TW Hya Association (TWA), but with our measured systemic  velocity, shows the binary is in fact not part of any known YMG.  This last system  also has lithium absorption in each component, and has strong and variable H$\alpha$ emission, setting an upper age limit of 15 Myr based on the lithium detection.

\end{abstract}


\section{Introduction}


Finding and characterizing young, low mass stars has been a prominent research field in recent years. There is interest in finding sub-stellar companions, especially planets, around young low-mass stars to provide observational constraints for planet formation \citep{CrockettSearchGiantPlanet2012, SpiegelSpectralPhotometricDiagnostics2012}.  As low-mass stars are the most common planet host \citep[e.g.,][]{dressing_occurrence_2013}, finding these stars provides targets for potential planet searches.  Additionally, characterizing young, low-mass stars is important to accurately infer the properties of any planets around them.   And as low-mass stars (late K and M stars) with masses less than $\sim$0.7 $M_{\odot}$ comprise more than 75\% of the stellar population \citep{reid_lowmass_1997, bochanski_luminosity_2010}, understanding the early evolutionary stages of low-mass stars is critical for understanding the stellar population as a whole \citep{DibTestinguniversalityIMF2014}. 

The first step to widening our understanding is by identifying these smaller --- and fainter --- young stars.   Young moving groups (YMG) are kinematically linked stars that formed together but have since dispersed \citep[e.g][]{zuckerman_young_2004}.  The properties of the group as a whole can be used to age-date the stars.  Over the past three decades, YMGs have been discovered based on kinematics with ages between $\sim$8 and 300 Myr old \citep{torres_search_2006}. In recent years, more low-mass members have been added to these groups by looking for kinematics and age indicators matched to known higher-mass members  \citep[e.g.,][]{montes_latetype_2001, schlieder_beta_2010, shkolnik_searching_2011, schlieder_likely_2012, shkolnik_identifying_2012, malo_banyan._2014-1,  gagne_banyan_2015, LiuHawaiiInfraredParallax2016, BowlerElusiveMajorityYoung2019}. Since we expect 3 of every 4 of the members to be M dwarfs, as in the rest of the galaxy, there are many more low-mass stars to be found.  Until they are, the census of the stellar population is incomplete.

In order to address the issue of incomplete stellar population for YMGs, the All-sky Co-moving Recovery Of Nearby Young Members (ACRONYM) search was created.  Thus far, the program has published 129 new low-mass members of the 40 Myr Tuc-Hor Moving Group \citep{kraus_stellar_2014}, 41 new low-mass members of the 23 Myr Beta Pic Moving Group \citep{ShkolnikAllskyComovingRecovery2017}, and 77 members to other groups \citep{SchneiderACRONYMIIIRadial2019}.  

\section{Candidate Selection}
The initial candidate selection process wass the same as that of \citet{ShkolnikAllskyComovingRecovery2017}, which is based identifying cool stars using  2MASS colors and magnitudes \citep{skrutskie_two_2006}, along with proper motions.   As youth is well-correlated with chromospheric activity, and high chromospheric activity can be measured photometrically in  the X-ray and ultraviolet (UV), surveys in the  X-ray with ROSAT \citep{voges_rosat_1999} and the UV with GALEX \citep{martin_galaxy_2005} were cross-referenced with our list of cool stars.

The coordinates and proper motions of stars from this list were also used as criteria. Certain combinations make it more likely for a star to be a kinematic fit into a YMG.  For each star, the full kinematics were calculated with different possible radial velocities and distances for that star, along with the known coordinates and proper motions for that star.  Stars where no plausible combination of distance and radial velocity allowed for membership were removed from our list, as in \citet{kraus_stellar_2014}. 

After selecting for UV or X-ray activity, we follow up with high-resolution spectroscopic observations to measure radial velocities and spectroscopic youth indicators.   Further follow up observations were carried out to search for radial velocity (RV) variability. In our activity-selected sample, we found three spectroscopic binaries (SBs) that also display signatures of youth (Table \ref{ystarprops}).  Binary stars are especially important to identify because they provide some of the best opportunities to calibrate stellar models and their predictions of fundamental stellar characteristics.  Measuring fundamental properties is crucial for understanding young, low-mass stars as they are not currently well modeled \citep{feiden_reevaluating_2012, SomersOlderColderImpact2015}.


\section{Targets}
 
Previous studies found evidence that all three of the spectroscopic binaries we analyze in this work are young. However, none of these objects had been identified as a binary system.

  \begin{deluxetable}{rlllrrrcrrrr}
 \tablewidth{0pt}
 \tabletypesize{\tiny}
 \tablecaption{Targets and their Parameters from Literature}
 \tablehead{
 \colhead{Object} & \colhead{RA} & \colhead{Dec} & \colhead{Spectral Type}  & \colhead{$\mu_{RA}$} & \colhead{$\mu_{Dec}$} & \colhead{V Mag} & \colhead{Distance}   &\colhead{FUV} &\colhead{NUV} & \colhead{HR1} & \colhead{HR2}
 \\
 \colhead{} & \colhead{} & \colhead{} & \colhead{} & \colhead{(mas/yr)} & \colhead{(mas/yr)} & \colhead{} & \colhead{(pc)} & \colhead{($\mu$Jy)} & \colhead{($\mu$Jy)} & \colhead{} & \colhead{}
  }
 \startdata 
J02543316-5108313 &02 54 33.16 & -51 08 31.4	&	M1.1	&	93.5	&	-11.8	&	12.1 & 44.3 & 21.5 & 79.4  & \nodata &	 \nodata	\\
J08355977-3042306 & 08 35 59.77 & -30 42 30.7	&	K4	&	-64.2	&	-14.5	&	11.3 & 62.9 & \nodata &	 \nodata & 0.15 & -0.20 \\
J10260210-4105537 &10 26 02.11 &-41 05 53.7	&	M0.5	&	-46.4	&	-1.8	&	12.6 & 84.9	& 16.5 & 113.6  & \nodata &	 \nodata\\
 \enddata
\tablecomments{ V magnitudes are from \citet{zacharias_fourth_2013}.  Composite spectral types are from \citet{kraus_stellar_2014} for J02543316-5108313, \citet{torres_search_2006} for J08355977-3042306, and \citet{riaz_identification_2006} for J10260210-4105537. Distances and proper motions are from Gaia \citep{BrownGaiaDataRelease2018}.  UV flux densities are from GALEX \citep{BianchiGALEXcatalogsUV2011}.  J08355977-3042306 was not observed with GALEX.  Hardness ratios (HR) are from ROSAT \citep{voges_rosat_1999}. \label{ystarprops} }
  \end{deluxetable}

\subsection{2MASS J02543316-5108313}
2MASS J02543316-5108313, also identified as GSC 08057-00342, was observed by \citet{torres_new_2000} in search for a new young moving group near the active young  star ER Eri.  Despite physical proximity and similar proper motions, 2MASS J02543316-5108313 was not classified into this new group called the Horologium association.   \citet{rodriguez_galex_2013} later determined there was  a $>$99\% chance that the star was in this association, now referred to as Tuc-Hor, based on its location and proper motion.  \citet{kraus_stellar_2014} measured the radial velocity as 13.8$\pm$0.4 km/s which confirmed that membership, but did not identify it as an SB2 as the lines were blended during the epoch of their observation.

This system was imaged by \citet{bergfors_lucky_2010} in a search for binaries, but no visual components were detected down to  0.1''.  This was followed up by \citet{janson_astralux_2012}, who classified it as a single star with X-ray emission.  \citet{krzesinski_mt._2012} measured a light curve of the system in search for variability, but no periodicity was detected nor were any flares.  Gaia DR2 \citep{BrownGaiaDataRelease2018} indicates that it has a proper motion companion with a 15" separation, which has an RV consistent with that of Tuc-Hor. \citet{ShanMultiplicityMDwarfsYoung2017} identified binaries with separations as low as 40 mas using MagAO, but did not detect another companion to this star that could potentially complicate the analysis of the system's kinematics. Its spectral type in the literature ranges from M1.1 \citep{kraus_stellar_2014} to M3 \citep{torres_new_2000}.    A low-resolution optical spectra from 2014 was presented in \citet{BowlerElusiveMajorityYoung2019}.  They found a spectral type of M0 and an H$\alpha$ EW of -3.2$\pm$0.3 \AA.

\subsection{2MASS J08355977-3042306}
2MASS J08355977-3042306, also referenced as CD-30 6530, was previously identified as a pre-main sequence K4Ve star by \citet{torres_search_2006}, who also measured lithium absorption with an equivalent width (EW) of 230 m\AA\ and an RV of 6.8 km/s. It was detected by ROSAT in the X-ray \citep{voges_rosat_1999}.  We observed it as part of a search for possible 23 Myr $\beta$PMG members \citep{ShkolnikAllskyComovingRecovery2017}.  

\subsection{2MASS J10260210-4105537}

We initially observed J10260210-4105537, an M0.5 star, due to two youth indicators indicators:  X-Ray emission and a strong H$\alpha$ EW of -6.9 \AA\ \citep{riaz_identification_2006}.  In a search for TWA members, \citet{rodriguez_new_2011} observed the system based on GALEX UV data.  Their high resolution spectrum showed lithium absorption, with an EW of 500$\pm$70 m\AA.  They also measured a spectral type of M1.  However, they did not identify the binarity of the system. A photometric survey by \citet{janson_astralux_2012} did not resolve the binary either and marked the system as a single star.  \citet{kiraga_asas_2012} analyzed the light curve from ASAS, which is a photometric survey with a resolution of 15" \citep{PojmanskiAllSkyAutomated2002}; they calculated a 0.42 day periodicty which they attributed to rotation and noted it was blended with a nearby star.  \citet{NaudPSYMWIDEsurveylargeseparation2017} looked for a potential companion at large separations and found none. \citet{GagneBANYANIXInitial2017} classify it as a likely contaminant to TWA,  and that its true membership is Lower Centaurus Crux. \citet{BowlerElusiveMajorityYoung2019} measure a spectral type M2 and a lithium EW of 410 m\AA\ based on low resolution spectroscopy; while they indicate that TWA is its most likely association, they acknowledge the need for a RV measurement.

\section{High Resolution Optical Spectra}
\subsection{Observations}
In order to confirm group membership, we measured a radial velocity and youth indicators by using high-resolution optical spectroscopy. Fourteen spectra in total were taken of the candidates between June 2009 and July 2017 from three different telescopes.  We aimed for a signal-to-noise of 25-30 at 7000 \AA\ (Table \ref{epochs}).  

\begin{figure}
\centering
\includegraphics[width=6.4in]{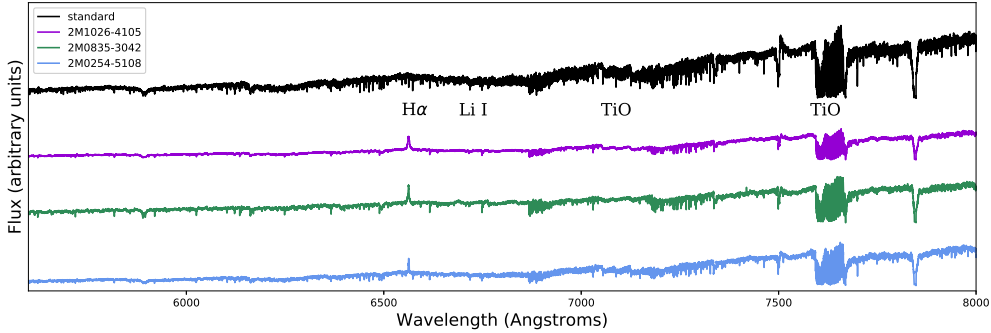}
\caption{Samples of the relevant portion of the spectra for our targets  taken with the echelle spectrograph on the du Pont telescope on December 11, 2015. The example standard star shown is GJ 908.  \label{spec}}
\end{figure}

Most of our spectra were taken on  the 2.5m Ir\'{e}n\'{e}e du Pont telescope at Las Campanas Observatories using the echelle spectrograph.\footnote{http://www.lco.cl/telescopes-information/irenee-du-pont/instruments/website/echelle-spectrograph-manuals/echelle-spectrograph-users-manual}  The detector is a SITe2k CCD with 24 $\mu$m pixels mounted at the Cassegrain focus.  We used a 0.75x4'' slit, which results in  a spectral resolution of $\approx$40,000 over 57 orders from 3625 \AA\ to 9325 \AA\ (Figure \ref{spec}).  We acquired ``milky'' flat field images at twilight every night, created by using a diffuser behind the slit, which distributes the light evenly across the detector.  We took Th-Ar spectra immediately following each observation in order to minimize variations from the spectrum. The images were reduced using the standard IRAF tasks \citep{TodyIRAFDataReduction1986}. 

 We obtained one spectrum with the CHIRON spectrograph  \citep{TokovininCHIRONFiberFed2013}, which is mounted on the Cerro Tololo Inter-American Observatory’s (CTIO) 1.5 m Small and Moderate Aperture Research Telescope System (SMARTS) 1.5 m telescope (NOAO Proposal ID: 2015A-0016; PI: B. Bowler).   This spectrum has 62 orders, with a spectral resolution of $\approx$28,000 and covers wavelengths between 4500 and 8900 \AA.  The spectrum was taken in queue mode.  The reduction, comprising a bias subtraction, flat fielding, order extraction, and wavelength calibration using Th-Ar, was done using the automated reduction pipeline.

For one target (2MASS J10260210-4105537), we obtained spectra taken on the Clay 6.5m telescope at Las Campanas Observatories, using the Magellan Inamori Kyocera Echelle (MIKE) spectrograph \citep{BernsteinMIKEDoubleEchelle2003}.  A 0.5'' slit was used to achieve a spectral resolution of $\approx$35,000 between 4900 and 10000 \AA.  These spectra were reduced using the facility pipeline \citep{kelson_optimal_2003}, which uses a milky flat to flat field the image, then extracts and calibrates the spectra.  Th-Ar spectra were used for wavelength calibration.  Additional adjustments to the wavelength calibration, typically less than 0.5 km/s, were made by using the telluric molecular oxygen A band between 7620 and 7660 \AA.

 \begin{deluxetable}{rrrrr}\label{epochs}
 \tabletypesize{\tiny}
 \tablewidth{0pt}
 \tablecaption{Spectroscopic Observations}
 \tablehead{
 \colhead{Object} & \colhead{UT Date} & \colhead{Telescope} & \colhead{Integration Time} & HJD
\vspace{-10pt}
 \\
 \colhead{} & \colhead{} & \colhead{} & \colhead{(s)} & \colhead{}
  }
 \startdata
J02543316-5108313 & 20090823 & du Pont & 600  & 2455066.91963418 \\
J02543316-5108313 & 20150622 & du Pont & 1350 &2457196.91753853   \\
J02543316-5108313 & 20151211 & du Pont & 750 & 2457367.60084198 \\
   \hline
J08355977-3042306 & 20140318 & du Pont & 400 & 2456744.59019459 \\
J08355977-3042306 & 20150112 & du Pont & 800 & 2457035.77692500  \\
J08355977-3042306 & 20150217 & SMARTS & 400 &  2457079.71270571  \\
J08355977-3042306 & 20151211 & du Pont & 1200 & 2457367.81581571  \\
J08355977-3042306 & 20170703 & du Pont & 1200 & 2457937.44805283 \\
   \hline
J10260210-4105537 & 20090614 & du Pont & 900 &  2454996.54189175 \\
J10260210-4105537 & 20091231 & Clay & 150 &     2455196.70946915 \\
J10260210-4105537 & 20101229 & Clay & 100 &     2455559.69428480 \\
J10260210-4105537 & 20110614 & Clay & 40  &     2455727.46221318 \\
J10260210-4105537 & 20151211 & du Pont & 1200 & 2457367.85578601  \\
J10260210-4105537 & 20151213 & du Pont & 900  & 2457369.83365365  \\
\enddata
 \end{deluxetable}

\subsection{Spectral Types and Masses}
 For calculating spectral indices, the blaze function was removed and the orders were combined to create one-dimensional spectra.  The IRAF task \textit{sbands} was used to calculate the indices used for spectral typing the targets.

For the MIKE data, we used the TiO5 index,  which, as defined by \citet{reid_palomarmsu_1995}, is the  ratio of the average flux from 7126 to 7135 \AA, covering the TiO  band, to the average continuum flux from 7042-7046 \AA.  This is then converted into a spectral type (SpT) by the empirically derived function:
\begin{eqnarray}
SpT=-10.775\times TiO5+8.2
\end{eqnarray}
where SpT is the M subclass (with -1 and -2 corresponding to K7 and K5 respectively) and TiO5 is the flux ratio.  It has an uncertainty of 0.5 subclasses.

 Since the spectra from the du Pont have the continuum of the index very close to the end of an order, we devised a variation of the TiO5 index for calculating spectral types of late K and early M stars for these spectra.  Instead of using the 7042-7046 \AA\ region for continuum, the region from 7558 to 7562 \AA\ --- which is also continuum \citep{kirkpatrick_standard_1991} --- is used.  The average flux from 7126 to 7135 \AA\ is divided by the average flux from 7558 to 7562 \AA\, and this ratio, henceforth known as TiO8, is converted into a spectral type by:
\begin{eqnarray}
SpT=5.3\times TiO8^2-15.2\times TiO8+7.2
\end{eqnarray}

\begin{figure}[htb!]
\centering
\includegraphics[width=3.4in]{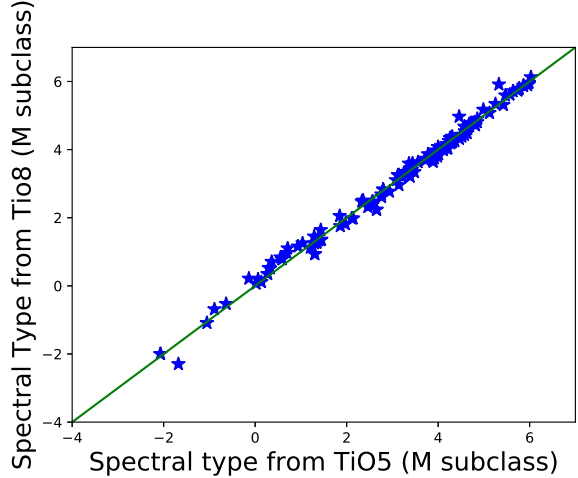}
\caption{A comparison of the TiO5 index with the TiO8 index for 141 low mass stars observed with MIKE. \label{tiov}}
\end{figure}

The resulting spectral types fit very closely with TiO5 for a sample of 141 MIKE spectra of low mass stars (Figure \ref{tiov}).  The residual standard error for this fit is 0.167, and combining this with the TiO5 uncertainty, the overall uncertainty for TiO8 is 0.5 spectral subclasses.  Due to intrinsic uncertainty, we round all spectral types to the nearest 0.5 subclass.

To calculate the spectral type of the individual components, we started with a grid of potential masses between 1.0 and 0.08 M$_\cdot$ in increments of 0.01 M$_\cdot$.  We then converted masses to spectral types using relations derived by \citet{pecaut_intrinsic_2013} and \citet{baraffe_new_2015}, interpolating for values not explicitly calculated for. We normalized both the masses and the spectral types to account for the fact that they do not have the same units.  We created a grid of component pairings that included masses and spectral types for each individual component.  We calculated the  composite spectral type weighted by I-band fluxes and mass ratio for each pairing.  The final component spectral types and masses that were the ones that minimized the differences between the measured composite spectral type and mass ratio simultaneously.



\subsection{Radial Velocities}\label{rvsec}
As the initial purpose for acquiring the spectra was to find new young moving group members by kinematics, we measure radial velocities for our stars, using the cross-correlation function (CCF) \citep{tonry_survey_1979}. The new SBs were identified by their  double-peaked CCFs.

To measure radial velocities of the individual components, we cross-correlated orders of the spectra with those of standard stars (Table \ref{tab:rvstd}) using the IRAF task \textit{fxcor} \citep{fitzpatrick_iraf_1993}.  We found the peak by fitting a Gaussian to the CCF and finding the center of that Gaussian.  We calculated uncertainties by combining the standard deviation of the velocities measured in each order with the uncertainty from wavelength calibration, calculated by the standard deviation from cross-correlating standards with other standards, and the uncertainty in the standards' RV of 0.15 km/s. The `A' component was taken to be the component with the higher CCF peak.  For spectra where the stars were especially blended, our own program created in LabVIEW was used to fit the sum of two Gaussians to the CCF.  This allowed us to have more control over the free parameters of the fit.

\begin{table}[htbp]
  \centering  
    \begin{tabular}{rrr}
    Star  & Spectral type & RV (km/s) \\
    \hline
 GJ 156 & K7 & 62.6 \\
GJ 273 & M3.5 & 18.3 \\
GJ 388 & M4.5 & 12.4 \\
GJ 406 & M4 & 19.5 \\
GJ 433 & M1.5 & 18.0 \\
GJ 514 & M0.5 & 14.6 \\
GJ 653 & K5 & 34.1		\\
GJ 699 & M4 & -110.5 \\
 GJ 908 & M1 & -71.1 \\

    \end{tabular}%
\caption{The RV standards used for cross-correlation. Radial velocities for the standards are from \citet{nidever_radial_2002} and have uncertainties of 0.15 km/s.  \label{tab:rvstd}}
\end{table}%

To calculate the systemic RV of these systems, we plotted the RV of the secondary component as a function of the RV of the primary component, as described in \citet{wilson_determination_1941}.   We then fit a straight line to the data using the \textit{scipy.odr} module \citep{VirtanenSciPyFundamentalAlgorithms2019}, which accounts for uncertainties in both dimensions. This  can then be used to calculate the systemic RV and mass ratio, $q$. 
 

\subsection{UVW and XYZ}
We calculated the three dimensional UVW space velocities, following  \citet{johnson_calculating_1987}, and the XYZ coordinates, following \citet{murray_transformation_1989} with errors propagated from the uncertainties in the RV measurements and distances.   The XYZ calculation requires a right ascension, declination, and distance; the UVW calculation requires those vlues in addition to proper motions and radial velocities.  Distances are calculated from Gaia DR2 parallaxes \citep{LuriGaiaDataRelease2018a, BrownGaiaDataRelease2018}.

\subsection{Youth Indicators}
There are various indicators with spectral and photometric data that can indicate a young age.  The three that we considered with these objects are lithium absorption, IR excess, and H$\alpha$ emission.
The presence of lithium absorption in a stellar spectrum is an indication of youth, because lithium is depleted via convective mixing early in a star's life \citep{bonsack_abundance_1959, bonsack_abundance_1960, skumanich_time_1972}.  The precise age at which this happens depends on spectral type, but for low-mass stars, which have large convective zones, this age is typically between 15-45 Myr \citep{chabrier_structure_1997}. 

Protoplanetary disks have lifetimes typically less than ten million years \citep[and references therein]{williams_protoplanetary_2011}.  Therefore, any star with such a protoplanetary disk must be young, identified by near- or mid-IR excess. Additionally, a stellar spectrum can reveal accretion signatures, such as strong and broadened H$\alpha$ emission \citep{barradoynavascues_empirical_2003}. H$\alpha$ emission can also be the result of chromospheric activity, as younger stars are  typically more active due to faster rotation periods  \citep[e.g.][]{skumanich_time_1972, linsky_stellar_1980, walter_xray_1988, soderblom_highresolution_1998, montes_library_1997}.  




\subsection{Binary Periods and Separations}
While a complete orbit is needed to measure orbital elements and physical properties, we can determine the upper limits of both the separation and the period of the binary system using our data.  The epoch in which the two components have the largest RV separation, $\Delta RV_{max}$, can be used with Kepler's 3rd Law and the total mass of the system to calculate the upper limit for the period, $P_{orb}$,  and semi-major axis, $a$.

\section{Optical Photometric Monitoring}
In addition to the high-resolution spectra, we obtained photometric observations to  measure a rotation period. We acquired  V band photometric data of 2MASS J08355977-3042306 from 2015 and 2017 taken with Lowell Observatory's 0.7-m telescope in robotic mode.  Two to three exposures of 70s each were taken at each visit to the field with uncertainties generally around 0.003 mag. The field was typically visited two times per night.  We have a total of 433 photometric observations.

Using the photometric data of 2MASS J08355977-3042306, we attempted to determine a rotation period of the stars by using a Lomb-Scargle periodogram \citep{Horneprescriptionperiodanalysis1986}.   We analyzed all the data together and also separated by the year they were taken.  

\section{Results and Analysis}
\subsection{2MASS J02543316-5108313}
\begin{figure}
\centering
\includegraphics[width=3.4in]{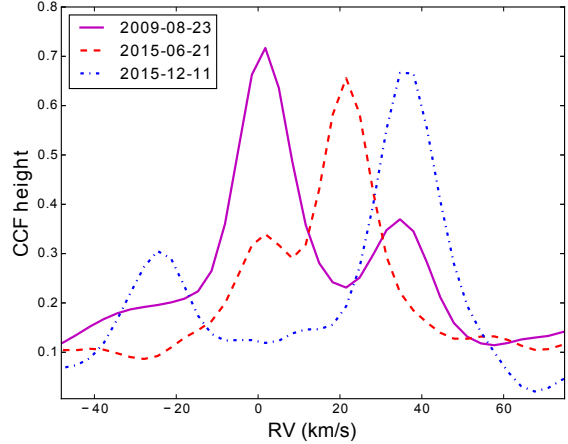}
\caption{CCFs for 2MASS J02543316-5108313 show binarity, with two peaks in each epoch. \label{ccf02}}
\end{figure}

We acquired three spectra of 2MASS J02543316-5108313 from the du Pont telescope, each of which show two peaks in the CCF (Figure \ref{ccf02}).  We measured the RV of each peak in all three epochs (Table \ref{02rv}).  

\begin{table}[htbp]
  \centering  
    \begin{tabular}{rrrr}
    UT Date  &   RV of A (km/s) & RV of B (km/s)\\
    \hline
20090823 &  1.0$\pm$0.2 & 30.4$\pm$0.7 \\
20150622 & 20.9$\pm$0.2 & 1.6$\pm$0.5 \\
20151211 &  36.7$\pm$0.7 & -22.5$\pm$0.9 \\
    \end{tabular}%
\caption{RVs for the two components of 2MASS J02543316-5108313. \label{02rv}}
\end{table}%

  \begin{figure}
\centering
\includegraphics[width=3.4in]{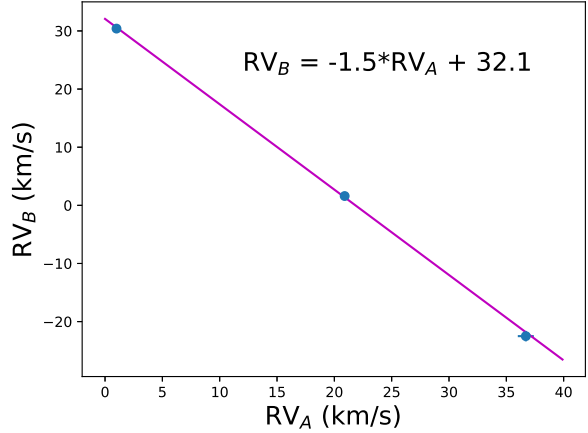}
\caption{The RV of component B plotted versus the RV of component A for the SB2 2MASS J02543316-5108313 from spectra collected over three epochs. \label{wilson02}}
\end{figure}
Based on our plot of the secondary's RV versus the primary's RV for this system (Figure \ref{wilson02}), we calculated a systemic RV of 13.0$\pm$0.1 km/s and a mass ratio of 0.68$\pm$0.03.   Based on the Gaia DR2 parallax \citep{BrownGaiaDataRelease2018}, the system at a distance of 43.8 pc.  With this distance, we then calculated UVW velocities (see Table \ref{kintable}) consistent with the Tuc-Hor members \citep{malo_banyan._2014-1}, which have an expected age of 40 Myr.

  \begin{figure}
\centering
\includegraphics[width=6.4in]{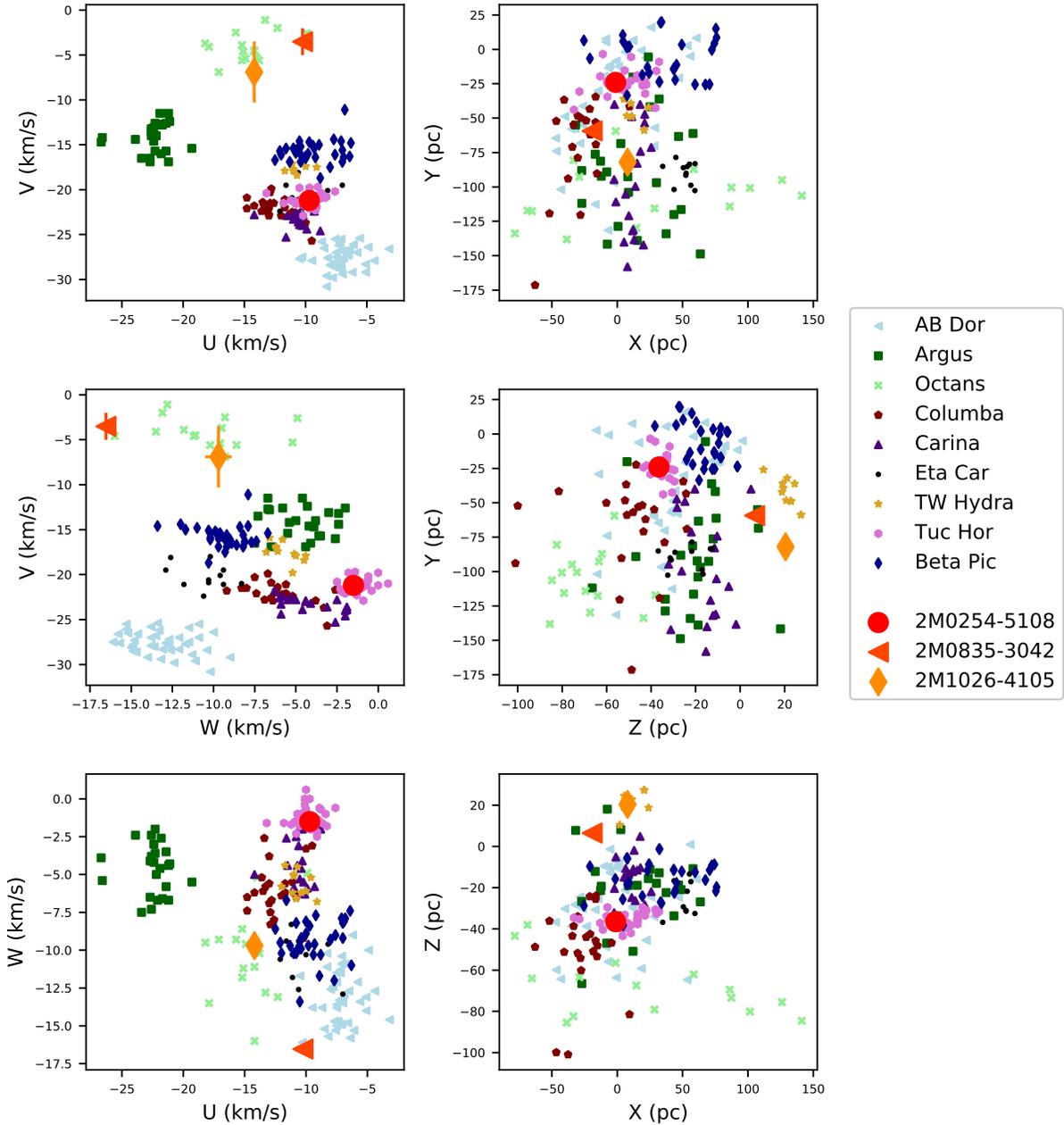}
\caption{UVW velocities and XYZ coordinates of our target and YMG members put 2MASS J02543316-5108313 in the Tuc-Hor moving group.  2MASS J08355977-3042306 and 2MASS J10260210-4105537 do not appear to be a member of any of these groups. \label{kin}}
\end{figure}

  \begin{deluxetable}{lrrrrrrrrr}
 \tablewidth{0pt}
 \tabletypesize{\tiny}
 \tablecaption{Kinematics and Orbital Parameters \label{kintable} }
 \tablehead{
 \colhead{Object} & \colhead{U} & \colhead{V} & \colhead{W}  & \colhead{X} & \colhead{Y} & \colhead{Z} & \colhead{P$_{orb}$} & \colhead{a} & \colhead{q} 
 \vspace{-10pt} \\
 \colhead{ } & \colhead{(km/s)} & \colhead{(km/s)} & \colhead{(km/s)}  & \colhead{(pc)} & \colhead{(pc)} & \colhead{(pc)} & \colhead{(days)} & \colhead{(AU)} & \colhead{} 
  }
 \startdata 
 2M0254-5108 & -10.0$\pm$0.1 & -21.6$\pm$0.1 & -1.2$\pm$0.1 & -1.2$\pm$0.1  & -24.7$\pm$1.7 & -37.6$\pm$2.8  & $<$32.1 & $<$0.17 & 0.68$\pm$0.03 \\
 2M0835-3042 & -9.9$\pm$0.5 & -3.6$\pm$1.5 & -15.8$\pm$0.2 & -18.4$\pm$1.0 & -56.4$\pm$3.1 & 6.2$\pm$0.3 & $<$134.0 & $<$0.56 & 0.59$\pm$0.06 \\  
 2M1026-4105 &   -14.2$\pm$0.4 & -6.79$\pm$3.4 & -9.7$\pm$0.9 & 7.9$\pm$0.1 & -82.0$\pm$1.0 & 20.4$\pm$0.2 & $<$9.4 & $<$0.08 &   0.59$\pm$0.30 \\
 \enddata
  \end{deluxetable}

Our measured composite spectral type for this system is an M2.2, which falls in the range of spectral types previously measured.  Using this, the mass ratio, and an estimated age of 40 Myr, we simultaneously fit for the accurate mass ratio and the composite spectral type to get the individual component masses and spectral types.  The primary component has a spectral type of M1.9, which we round to M2.0, and a mass of 0.40 $M_{\odot}$, while the secondary has a spectral type of M2.9, which we round to M3.0, and a mass of 0.27 $M_{\odot}$. The lack of lithium from either component puts an additional lower age limit of 20 Myr on the system \citep{baraffe_new_2015}.  The upper limits of the period and the semi-major axis are 32.1 days and 0.17 AU, respectively.

\subsection{2MASS J08355977-3042306}

\begin{figure}
\centering
\includegraphics[width=3.4in]{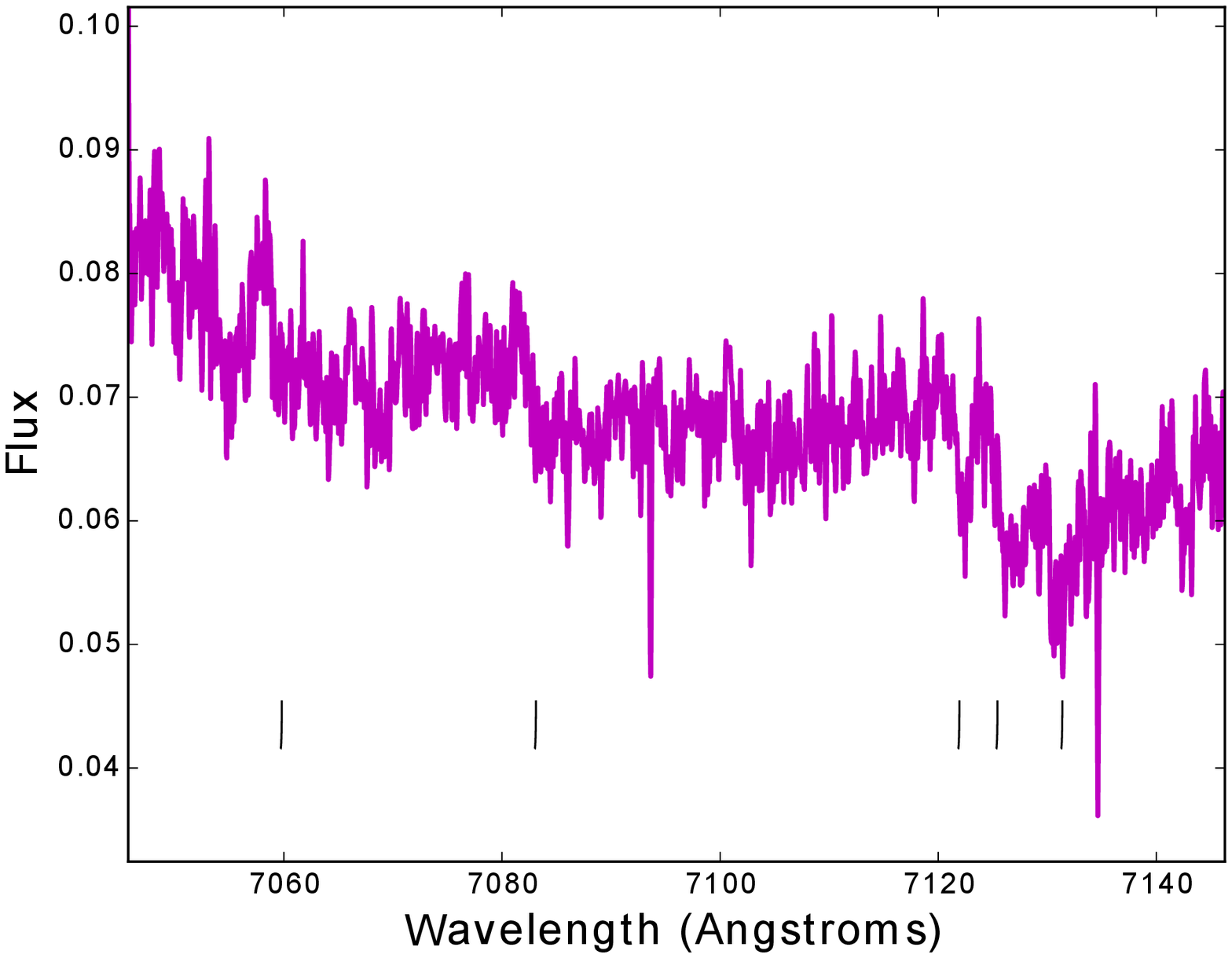}
\caption{TiO absorption bands in the spectrum of 2MASS J08355977-3042306 (taken with the echelle spectrograph on the du Pont telescope on March 18, 2014) indicates that the composite spectral type is K5 or later. \label{tio08}}
\end{figure}

We used the TiO8 index and measured a composite spectral type of K7.1 for 2MASS J08355977-3042306. This is a later spectral type than the previously measured one of K4 \citep{torres_search_2006}.  However, the star has TiO absorption bands (Figure \ref{tio08}), which appear only in stars with spectral types of K5 or later \citep{reid_palomarmsu_1995}, thus ruling out K4 as a possibility.  

\begin{figure}
\centering
\includegraphics[width=3.4in]{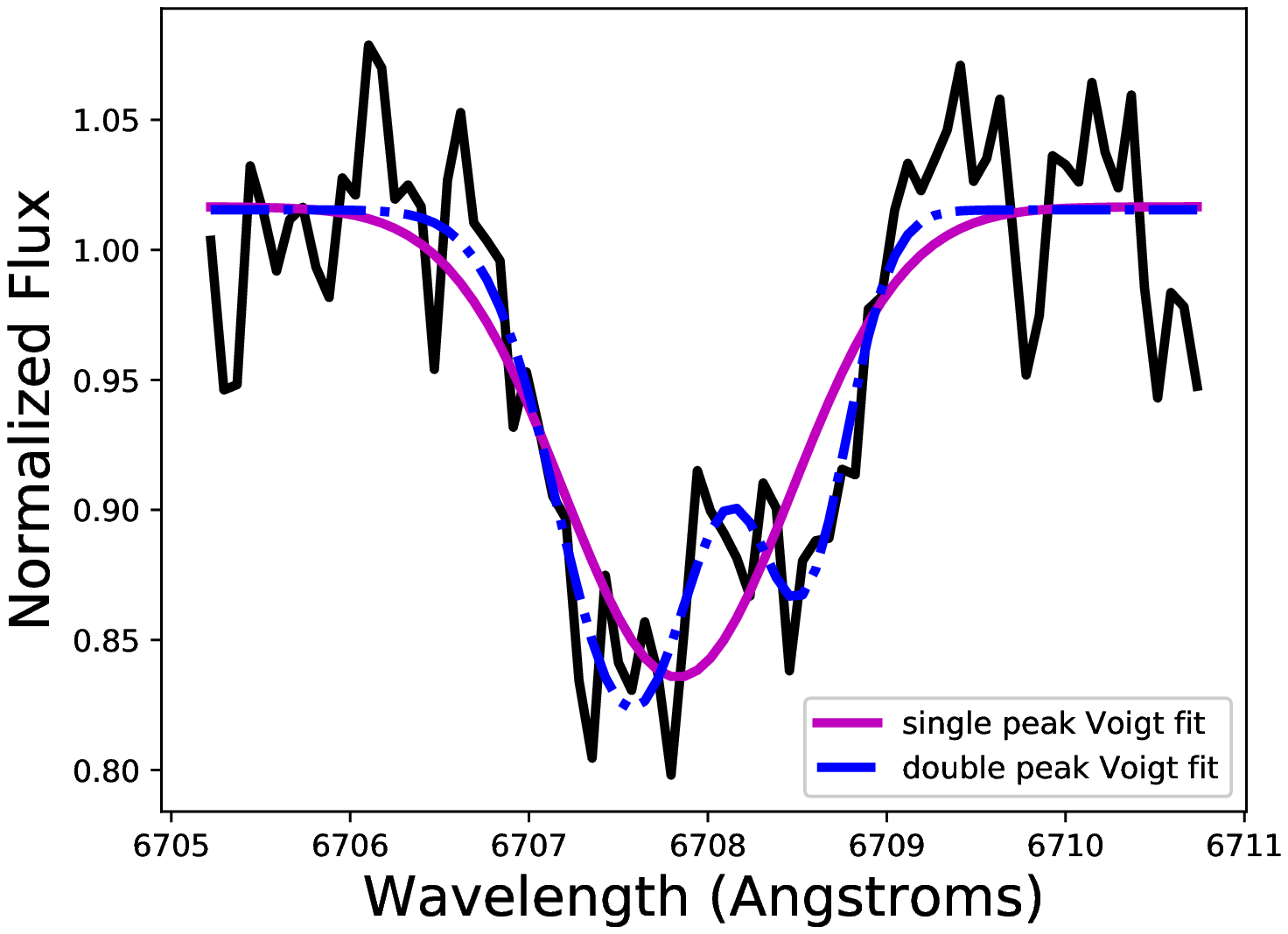}
\caption{Models of the lithium absorption at 6708 \AA\ for 2MASS J08355977-3042306 from the March 18, 2014 spectrum  comparing the fit of a single peak versus a double peak.  The double peak fits better, even considering the additional free parameters. \label{fit:lith}}
\end{figure}

The lithium absorption line at 6708 \AA\ (Figure \ref{fit:lith}) has a combined equivalent width of 0.25$\pm$0.10 \AA\, with the presence of lithium in both objects spectroscopically resolved.  


%

  \begin{figure}
\centering
\includegraphics[width=3.4in]{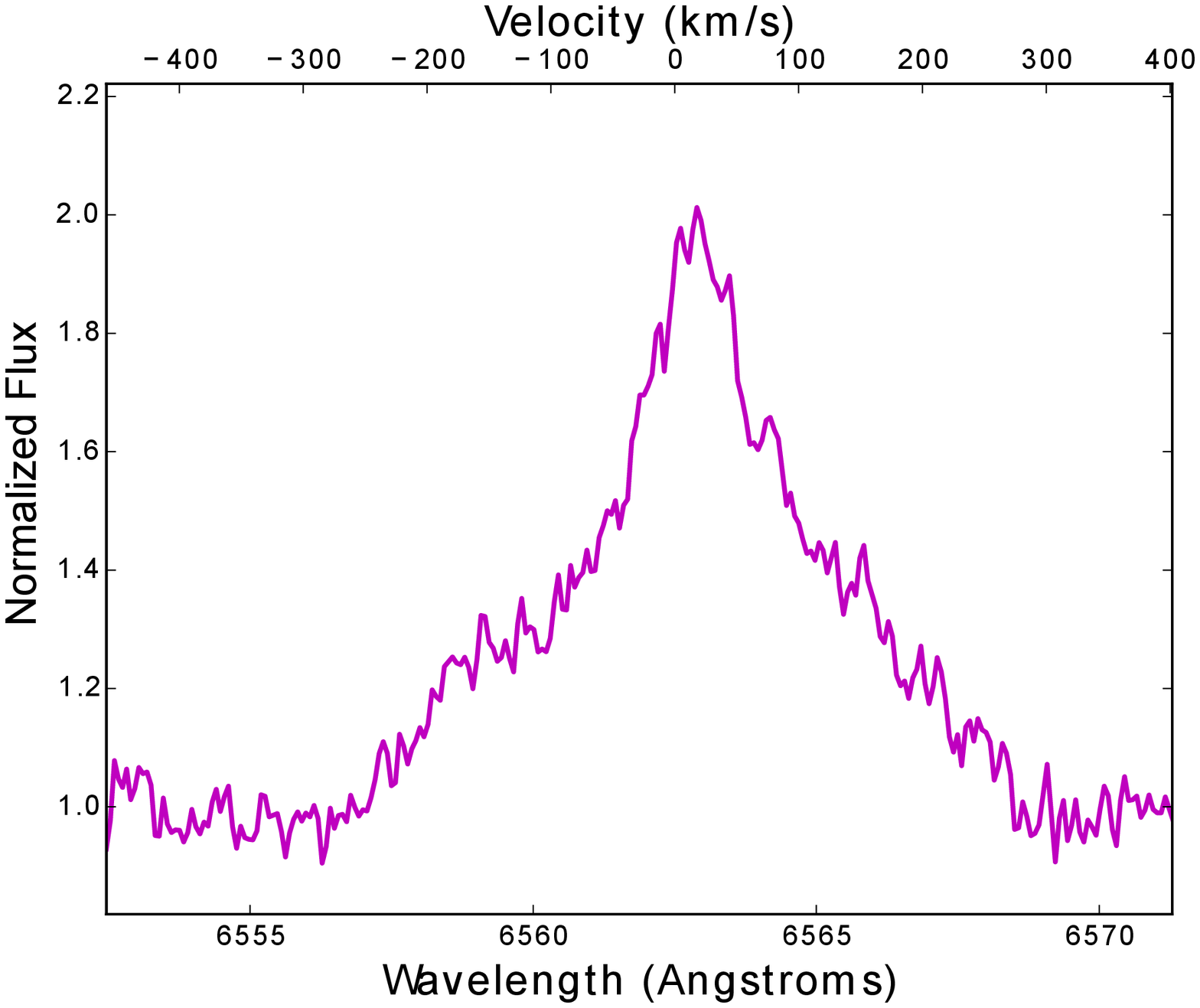}
\caption{H$\alpha$ emission for 2MASS J08355977-3042306 from March 18, 2014.  The 10\%-velocity width is $\sim$350 km/s, which is typically indicative of accretion. \label{ha08}}
\end{figure}

We measured the EW of the H$\alpha$ line at 6562.8 \AA\ to be -3.0$\pm$0.3 \AA, in agreement the value of -2.9 \AA\ measured by \citet{torres_search_2006}, using the standard convention that emission features have negative EWs.  We also measured the 10\%-velocity width, defined as the full width of the line at 10\% of the height (see Figure \ref{ha08}), to be $\sim$350 km/s.  This width is indicative of accretion \citep{white_very_2003}.  However, it should be noted that there is no obvious IR excess based on its 2MASS and WISE  colors \citep{wright_wide-field_2010}, which implies the system lacks warm dust  \citep[e.g][]{spangler_dusty_2001}.  It is possible that the system is accreting gas, which would not appear as IR excess \citep{hoadley_evolution_2015}.

To confirm 2MASS J08355977-3042306 is a binary and not a single star, we fit the lithium line with both a single peak and a double peak function using both Gaussian and Voigt models (see Figure \ref{fit:lith}).  For both models,  Bayesian Information Criteria  \citep{KassBayesfactors1995}  --- which takes into account the number of free parameters ---  indicated that the double peak model clearly fit better, with a difference of 14.9 for the Gaussian and 10.6 for the Voight.

\begin{figure}
\centering
\includegraphics[width=6.4in]{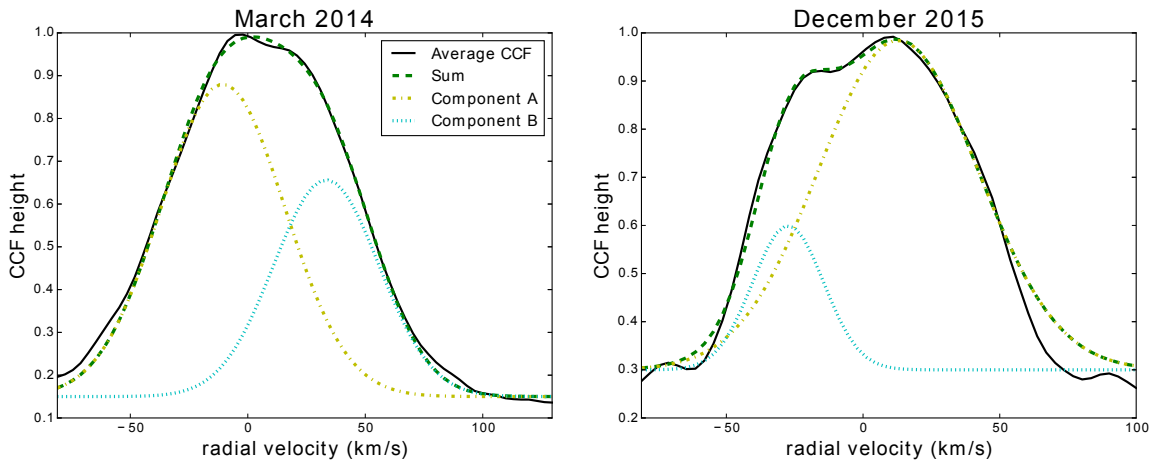}
\caption{Deblended CCFs for 2MASS J08355977-3042306 on March 28, 2014 (left) and December 11, 2015 (right). \label{deblend08}}
\end{figure}

\begin{figure}[!htb]
\centering
\includegraphics[width=3.4in]{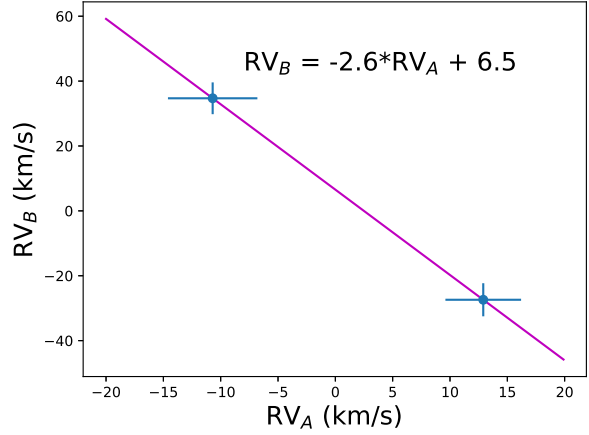}
\caption{The RV of component B as a function of the RV of component A for 2MASS J08355977-3042306. \label{wilson08}}
\end{figure}

For two of the epochs, the CCF could be deblended into separate peaks (Figure \ref{deblend08}).  We measured the RVs of those peaks (Table \ref{08rv}), and then plotted our measurements (Figure \ref{wilson08}).  The resulting systemic RV is 4.7$\pm$1.6 km/s; the mass ratio is 0.59$\pm$0.06.  Based on this mass ratio, the primary has a mass of 0.79 $M_{\odot}$, which corresponds to a spectral type of K5.4, which we round to K5.5, and the secondary has a mass of 0.46 $M_{\odot}$, which corresponds to a spectral type of M1.4, which rounds to M1.5. As lithium is depleted within 15 Myr for 0.4-0.5 $M_{\odot}$, stars \citep{baraffe_new_2015}, the system must be younger than that.  The upper limit of the period is 134.0 days; the upper limit of the semi-major axis is 0.56 AU.

\begin{table}[htb]
  \centering  
    \begin{tabular}{rrrr}
    UT Date  &   RV of A (km/s) & RV of B (km/s)\\
    \hline
March 28, 2014 &  -10.7$\pm$3.9 & 34.7$\pm$4.9 \\
December 11, 2015 & 12.9$\pm$3.3 & -27.4$\pm$5.1 \\
    \end{tabular}%
\caption{RVs for 2MASS J08355977-3042306. \label{08rv}}
\end{table}%

\begin{figure}[!htb]
\centering
\includegraphics[width=3.4in]{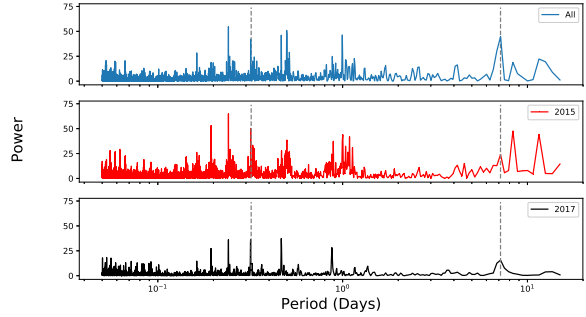}
\caption{Periodograms for 2MASS J08355977-3042306 of the photometric data for all the data (top), just the 2015 data (middle), and just the 2017 data (bottom).  Peaks at 0.32 and 7.15 days occur in all three.  \label{periodogram}}
\end{figure}

The periodograms we calculated (Figure \ref{periodogram}) all show periodicity at 0.32 days, around 1 day, and at 7.15 days.  The peaks at 1 day are likely to be caused by the nature of sampling the data every night.  The peaks at 0.32 days could be the rotation period of the primary.  However, this would put its velocity close to the break up speed.  The peaks at 7.15 days may be a more physically realistic estimate of the rotation period, and thus the orbital period. Both the 0.32 day and the 7.15 day periods are within the range of rotation periods common for stars of this age \citep{RebullRotationLowmassStars2018}.  Simulations suggest that tidal locking takes at minimum 15 Myr \citep{FlemingRotationPeriodEvolution2019}, so we would not expect this to be reflective of the orbital period.

 A parallax measurement from Gaia \citep{GaiaCollaborationGaiamission2016} indicates a distance of  62.7$\pm$1.6 pc. Using this distance and the systemic RV, we calculate its UVW velocities as (-9.9$\pm$0.5,  -3.6$\pm$1.5,  -15.8$\pm$0.2) km/s and XYZ coordinates as (-18.4$\pm$1.0, -56.4$\pm$3.1, 6.2$\pm$0.3) pc.

Plotting these with the locations of YMG in both 3D velocity space and positional space (Figure \ref{kin}), the system is not consistent with any known YMG. There are three likely possibilities for that; we explore each of those three options below.


\begin{itemize}
\item If the system formed in a known YMG, it could have been ejected after formation due to three-body gravitational interactions \citep{sterzik_escape_1995}.  However, the initial interaction would produce a star with higher velocities ($\gtrsim$40 km/s) than 2MASS J08355977-3042306 has, so while we cannot rule out this option completely, it is the least likely scenario of the three \citep{perets_properties_2012}.

\item While most stars form in clusters \citep[and references therein]{evans_physical_1999}, stars and binaries are capable of forming  in isolated environments \citep{adams_modes_2001}.  Based on the spatial distribution of young stellar objects, at least 10\% of low-mass stars appear to form in isolation \citep{bressert_spatial_2010}.  In this case, it would not have kinematics that matched with any other star's. 
 
\item  Surveys of low-mass stars are only complete to a distance of 25 pc \citep{shkolnik_identifying_2012}. Most stars in known YMGs are within 90 pc of the Sun  \citep{malo_banyan._2014-1}. It is very possible that 2MASS J08355977-3042306 could be the first star in a yet-to-be-discovered YMG, much like TW Hydra once was.  
\end{itemize}

Since it does not fit into any YMG, we questioned whether the system was truly young.  RS CVn systems are close, active binaries, where at least one of the components is an evolved star \citep{hall_rs_1976}.  They have lithium and large 10\% velocity widths \citep{eker_h-alpha_1987, pallavicini_lithium_1992, bopp_extremely_1993, montes_excess_1995}.  However, while RS CVn systems have lithium, their EWs are less than 100 m\AA\ which is much less than that of the 2MASS J08355977-3042306 \citep{neuhaeuser_optical_1997}.  Additionally, all currently identified RS CVn stars have spectral types earlier than that of 2MASS J08355977-3042306.

\subsection{2MASS J10260210-4105537}

\begin{figure}
\centering
\includegraphics[width=3.4in]{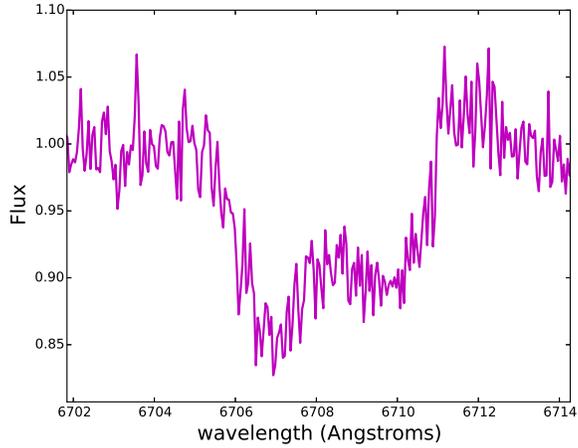}
\caption{Lithium absorption at 6708 \AA\ for 2MASS J10260210-4105537, with both components resolved. The presence of lithium puts an upper age limit on the system of 15 Myr. \label{li10}}
\end{figure}

We acquired data in six epochs for this star --- three from the du Pont and three from Magellan.  All six epochs show lithium absorption of both components resolved (Figure \ref{li10}).  We measured a composite spectral type of M1.5$\pm$0.5 for this system.   The absorption features are wide and shallow, indicating that both components are relatively rapid rotators, which is common in young stars.

\begin{figure}
\centering
\includegraphics[width=3.4in]{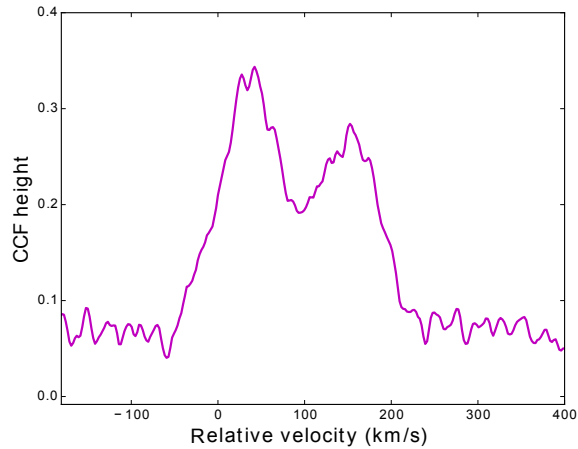}
\caption{CCF for 2MASS J10260210-4105537 from December 2009 using GJ 908 as a template. The radial velocity of both components could be measured in this epoch. \label{ccf10}}
\end{figure}
\begin{figure}
\centering
\includegraphics[width=3.4in]{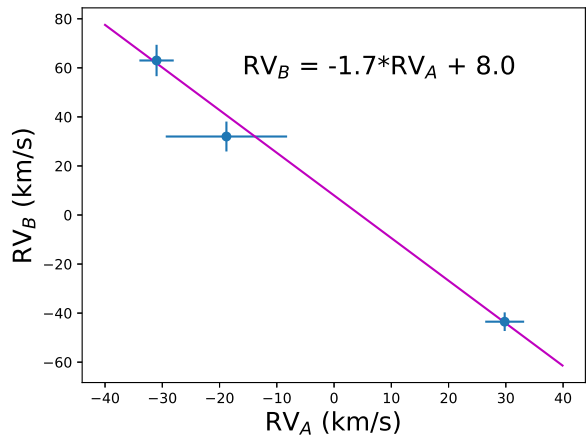}
\caption{The RV of component B as a function of the RV of component A for 2MASS J10260210-4105537. From this, we calculated a mass ratio of 0.59$\pm$0.30 and a systemic RV of 3.0$\pm$3.5 km/s \label{wilson10}}
\end{figure}

Five of the CCFs showed displaced velocity-resolved components (Figure \ref{ccf10}).  However, we were only able to calculate RVs from the CCF for three of the six epochs (Table \ref{10rv}), because we were unable to deblend the others due to poor S/N and lack of velocity separation between the components.   After fitting a line to the data in the RV plot (Figure \ref{wilson10}), we calculated a mass ratio of 0.59$\pm$0.30, with the primary at a mass of 0.50 $M_{\odot}$ and a spectral type of M0.9, which rounds to M1.0, and the secondary at a mass of 0.29 $M_{\odot}$ and a spectral type of M3.0. For stars with a mass of 0.5 $M_{\odot}$, lithium is typically depleted by 15 Myr \citep{baraffe_new_2015}, putting an upper age limit on the system.  The systemic RV is 3.0$\pm$3.5 km/s.  The large uncertainties are due to the broad CCF peaks. We calculate an upper limit of the period to 9.4 days and the upper limit of the semi-major axis to be 0.08 AU.

\begin{table}[htb]
  \centering  
    \begin{tabular}{rrrr}
    UT Date  &   RV of A (km/s) & RV of B (km/s)\\
    \hline
		June 14, 2009 &  -31.0$\pm$3.0 & 63.0$\pm$6.4 \\
		December 31, 2009 &  29.8$\pm$3.4 & -43.5$\pm$3.8 \\
December 11, 2015 & -18.8$\pm$10.6 & 32.0$\pm$6.1 \\
    \end{tabular}%
\caption{RVs for J10260210-4105537. \label{10rv}}
\end{table}%

  Using this the distance of 84.9$\pm$1.0 pc from Gaia DR2 along with our RV, we calculated UVW to be ( -14.2$\pm$0.4, -6.79$\pm$3.4,  -9.7$\pm$0.9) km/s and XYZ to be ( 7.9$\pm$0.1, -82.0$\pm$1.0, 20.4$\pm$0.2) pc. Based on both UVW and XYZ, the system does not appear to be a match for any YMG (Figure \ref{kin}) or nearby star forming regions \citep{mamajek_uv_2015}. While it fits well in Octans ($\approx$20 Myr) in UVW space, it does not match in XYZ space. This young binary is another relatively distant, isolated, young system.

\begin{figure}[htb!]
\centering
\includegraphics[width=3.4in]{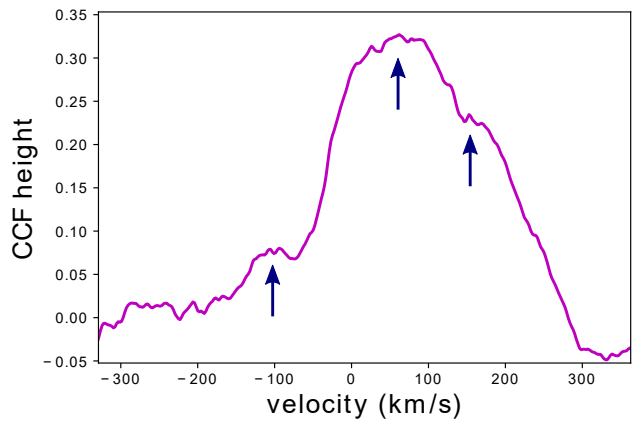}
\caption{CCF from December 2010 shows indications of three stars. \label{threepeaks}}
\end{figure}

However, there are indications in the spectra and CCF, particularly the one from December 2010 (Figure \ref{threepeaks}), that this system may in reality be a triple system.  While we believe this to be fairly unlikely as evidence is not seen in several other epochs, if true, the calculated RV and UVW velocities would not be accurate as they are calculated based on the gravitational interactions between two bodies.  In that case, the system may fit into a YMG. 

\subsubsection{Spectral Features and Flare Activity}

H$\alpha$ is  in emission in all of the spectra for 2MASS J10260210-4105537.  The equivalent width of H$\alpha$ ranged from -8 to -10 \AA\ in the first five epochs.  These values make it unlikely that the star is still accreting \citep{white_very_2003}.  

  \begin{figure}[htb!]
\centering
\includegraphics[width=3.4in]{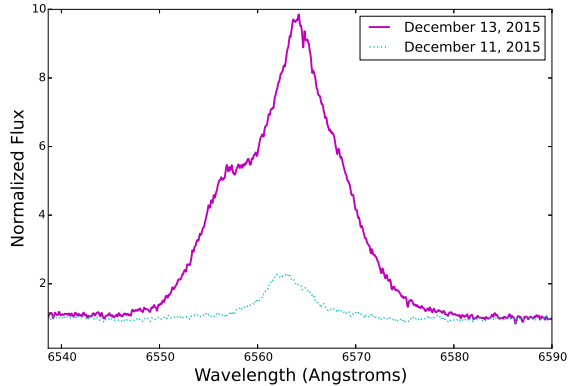}
\caption{H$\alpha$ emission for 2MASS J10260210-4105537 increases by an order of magnitude in two days. \label{ha10}}
\end{figure}

\begin{figure}[htb!]
\centering
\includegraphics[width=3.4in]{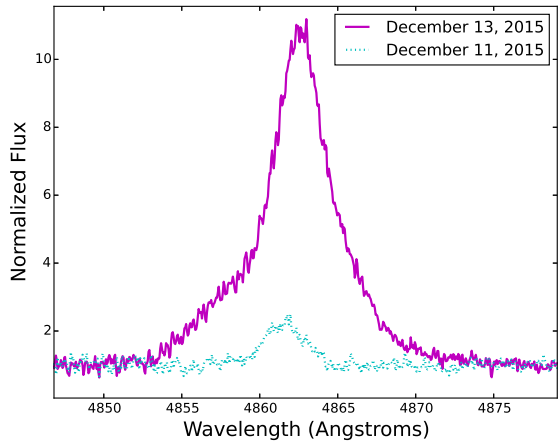}
\caption{H$\beta$ emission for 2MASS J10260210-4105537 increases by over an order of magnitude in two days due to chromospheric activity. \label{hb10}}
\end{figure}

\begin{figure}[htb!]
\centering
\includegraphics[width=6.4in]{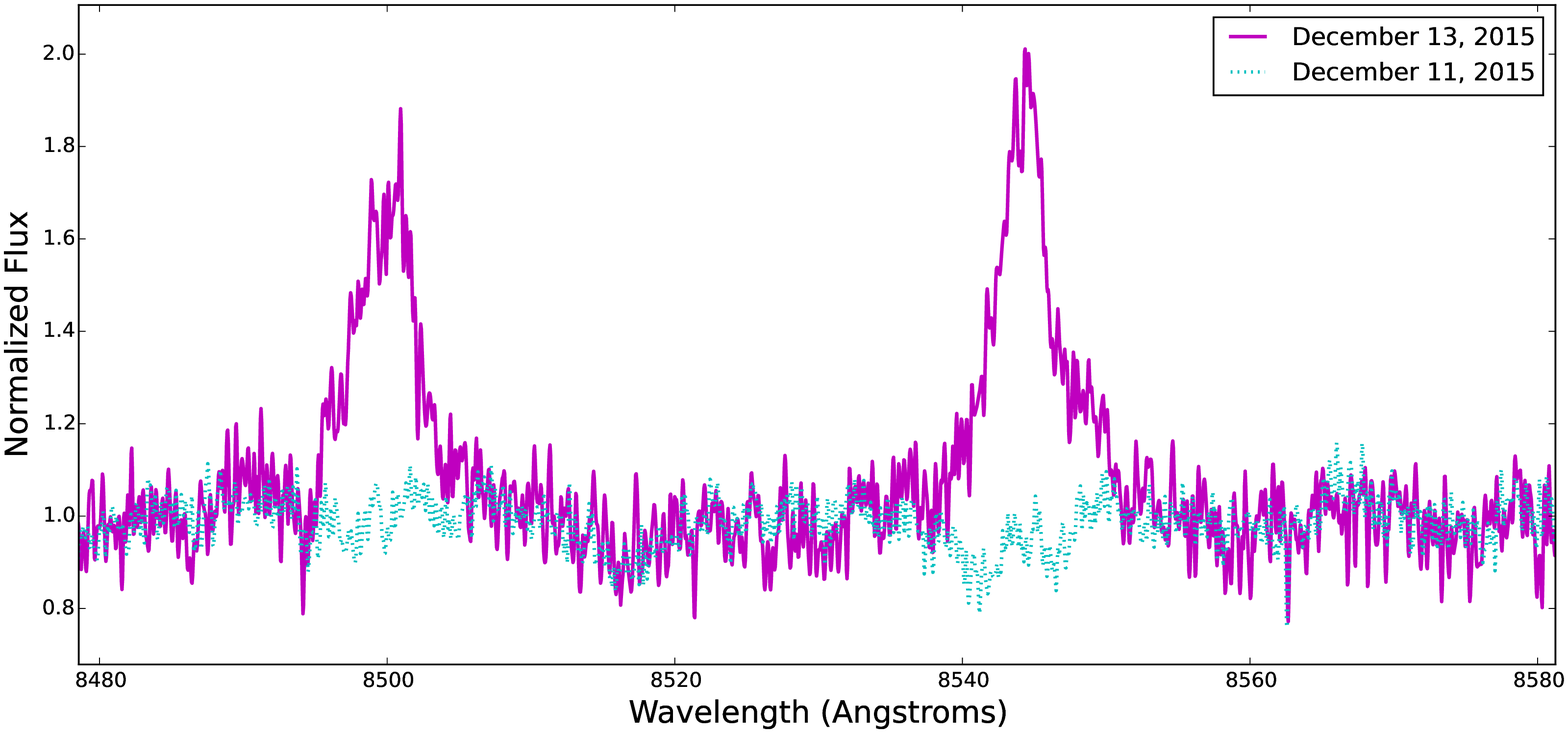}
\caption{Two lines of the Ca II triplet for 2MASS J10260210-4105537 were in absorption on December 11, 2015, but emission on December 13, 2015. \label{catrip}}
\end{figure}

However, in the final epoch from December 13, 2015, which was taken only two days after the previous one, its equivalent width was -92 \AA\ (Figure \ref{ha10}).  Comparable increases occurred in all of the hydrogen Balmer series features (Figure \ref{hb10}) and  helium features at 4471, 4921, 5018, 5876, and 6678 \AA.  Additionally, other features that were in absorption on December 11, 2015 were in emission on December 13, 2015, such as Fe II at 5169 \AA, Fe I at 5328 \AA, Na I doublet at 5890 and 5896 \AA, O I at 7773 \AA, and the Ca II triplet (Figure \ref{catrip}) at 8498, 8542, and 8662 \AA.

 The cause of activity can not be determined from the data we have now.  Possible causes include regular activity from one or both stars typically seen in young, low-mass stars \citep{noyes_rotation_1984}, although generally not at this level.  It could also be due to the  interactions between the binaries \citep{brandner_multiplicity_1996}.  And while it does not appear to be actively accreting and there is no IR excess, there could still be brief accretion events.  Further observations are needed to determine the frequency and magnitude of these events to have a better understanding of the cause.
	
	\section{Conclusions}

In this paper, we present three new young, low-mass SBs discovered with high-resolution optical spectra.  Using the component RVs we measured for each epoch, we calculated mass ratios and the systemic RVs for all three SBs.   Using the systemic RVs with their coordinates and proper motions, we calculated full 3D  velocities to search for potential membership of known YMGs. We also determined the masses and spectral types for the individual components.  For 2MASS J02543316-5108313 our RV measurements confirmed the Tuc-Hor membership from \citet{kraus_stellar_2014}.   All characteristics, including the lack of lithium, are consistent with Tuc-Hor's age of 40 Myr. 2MASS J08355977-3042306 is an SB2 with lithium in both components.  This puts an upper age limit of 15 Myr on the system.  Despite this, it does not fit kinematically with any known YMG. 2MASS J10260210-4105537 was previously known to be young based on its lithium absorption and was a potential TWA member.  However, it too does not fit with any known moving group based on its kinematics.  It is possible that these two ``homeless systems'' are the first stars to be discovered in new YMGs that are $<$15 Myr old and only $\approx$60-100 pc away.  Discovering new, nearby YMGs associated with these two young SBs, as has been done for other stars post-Gaia \citep[e.g.,][]{FahertyNewKnownMoving2018}, would be valuable for developing a more complete understanding of the evolution of low-mass stars.  






\section*{Acknowledgements}
This work was supported by the NASA/Habitable Worlds grant NNX16AB62G (PI E. Shkolnik). B.P.B. acknowledges support from the National Science Foundation grant AST-1909209.   L.F. would like to thank Lisa Prato for many helpful discussions and Joleen Carlberg for patiently guiding me though the spectroscopic data reduction process for the first time.  This research has made use of the VizieR catalogue access tool, CDS, Strasbourg, France. The original description of the VizieR service was published in A\&AS 143, 23.  This research has made use of the SIMBAD database,operated at CDS, Strasbourg, France  \citep{WengerSIMBADastronomicaldatabase2000}. This research has made use of the NASA/ IPAC Infrared Science Archive, which is operated by the Jet Propulsion Laboratory, California Institute of Technology, under contract with the National Aeronautics and Space Administration. This work has made use of data from the European Space Agency (ESA) mission {\it Gaia} (\url{https://www.cosmos.esa.int/gaia}), processed by the {\it Gaia} Data Processing and Analysis Consortium (DPAC, \url{https://www.cosmos.esa.int/web/gaia/dpac/consortium}). Funding for the DPAC has been provided by national institutions, in particular the institutions participating in the {\it Gaia} Multilateral Agreement. 

\bibliography{08-30}
\end{document}